\documentclass[twocolumn,prd,showpacs,floatfix]{revtex4}
\usepackage{epsfig}

\newcommand{\be}{\begin{equation}}
\newcommand{\ee}{\end{equation}}  
\newcommand{\bea}{\begin{eqnarray}}
\newcommand{\eea}{\end{eqnarray}}  
\newcommand{\gag}{g_{a\gamma}}

\begin{document}

\title{Photon-axion conversion as a mechanism for supernova dimming:\\
Limits from CMB  spectral distortion}

\author{Alessandro Mirizzi$^{1,2}$,
Georg G.~Raffelt$^{1}$,
and Pasquale D.~Serpico$^{1}$}

\affiliation{
$^1$Max-Planck-Institut f\"ur Physik (Werner-Heisenberg-Institut),
F\"ohringer Ring 6, 80805 M\"unchen, Germany\\
$^2$Dipartimento di Fisica and Sezione INFN di Bari,
                Via Amendola 173, 70126 Bari, Italy}

\date{\today}

%%%%%%%%%%%%%%%%%%%%%%%%%%%%%%%%%%%%%%%%%%%%%%%%%%%%%%%%%%%%%%%%%%%%%%
\begin{abstract}
%%%%%%%%%%%%%%%%%%%%%%%%%%%%%%%%%%%%%%%%%%%%%%%%%%%%%%%%%%%%%%%%%%%%%%
  Axion-photon conversion induced by intergalactic magnetic fields has
  been proposed as an explanation for the dimming of distant
  supernovae of type Ia (SNe~Ia) without cosmic acceleration.  The
  effect depends on the intergalactic electron density $n_e$ as well
  as the $B$-field strength and domain size.  We show that for
  $n_e\alt 10^{-9}~{\rm cm}^{-3}$ the same mechanism would cause
  excessive spectral distortion of the cosmic microwave background
  (CMB).  This small-$n_e$ parameter region had been left open by the
  most restrictive previous constraints based on the
  dispersion of quasar (QSO) spectra. The combination of CMB and QSO
  limits suggests that the photon-axion conversion mechanism can only
  play a subleading role for SN~Ia dimming.  A combined analysis of all
  the observables affected by the photon-axion oscillations would be
  required to give a final verdict on the viability of this model.
\end{abstract}
%%%%%%%%%%%%%%%%%%%%%%%%%%%%%%%%%%%%%%%%%%%%%%%%%%%%%%%%%%%%%%%%%%%%%%
\pacs{98.80.Es, % Observational cosmology
98.80.Cq,
 %Particle-theory and field-theory models of the early Universe 
14.80.Mz.
 %Axions and other Nambu-Goldstone bosons (Majorons, familons, etc.)
\hfill Preprint MPP-2005-53}

\maketitle

%%%%%%%%%%%%%%%%%%%%%%%%%%%%%%%%%%%%%%%%%%%%%%%%%%%%%%%%%%%%%%%%%%%%%%
\section{Introduction}
%%%%%%%%%%%%%%%%%%%%%%%%%%%%%%%%%%%%%%%%%%%%%%%%%%%%%%%%%%%%%%%%%%%%%%

Supernovae of type Ia (SNe~Ia) at redshifts $0.3 \lesssim z \lesssim
1.7$ appear fainter than expected from the luminosity-redshift
relation in a decelerating
Universe~\cite{supnovR,supnovP,Riess:2004nr}.  On the other hand, the
cosmic microwave background (CMB) anisotropy and large-scale structure
observations suggest that the Universe is spatially flat, with a
matter density of approximately 30\% of the critical
density~\cite{Spergel:2003cb,Tegmark:2003ud}.  The ``concordance
model'' thus implies that the Universe must be accelerating today
because it is dominated by a ``dark energy'' component (about 70\% of
the critical density) with an equation of state $w=p/\rho\approx -1$.

The lack of a satisfactory fundamental explanation for this component
has triggered wide-ranging theoretical investigations of more or less
exotic scenarios~\cite{Carroll:2003qq}.  Some years ago Cs\'aki,
Kaloper and Terning~\cite{Csaki:2001yk} (CKT~I) suggested that the
observed achromatic dimming of distant SNe~Ia may be a consequence of
the mixing of photons with very light and weakly coupled axion-like
particles in the intergalactic magnetic fields. Though still requiring
some non-standard fluid (e.g.~with $p/\rho\simeq -1/3$) to fit the
flatness of the universe, this model seemed capable to explain the
SN dimming through a completely different mechanism without
apparently affecting other cosmological observations.

Later it was recognized that the conclusions of CKT~I can be
significantly modified when the effects of the intergalactic plasma on
the photon-axion oscillations are taken into
account~\cite{Deffayet:2001pc}.  Assuming an electron density
$n_e\approx n_{\rm baryons}= n_{\gamma}\eta \sim 10^{-7}{\rm
cm}^{-3}$, the model is ruled out in most of the parameter space
because of either an excessive photon conversion or a chromaticity of
the dimming. Only fine-tuned parameters for the statistical properties
of the extragalactic magnetic fields would still allow this
explanation.  On the other hand, Cs\'aki, Kaloper and
Terning~\cite{Csaki:2001jk} (CKT~II) criticized the assumed value of
$n_e$ as being far too large for most of the intergalactic
space, invoking observational hints for a value
at least one order of magnitude smaller. For $n_e\alt 2.5\times
10^{-8}$~cm$^{-3}$, the photon-axion mixing hypothesis works even
better with the plasma, because the constraints from CMB anisotropies
via photon-axion conversion can be relaxed.

If distant SNe~Ia are dimmed by this mechanism, the same would apply
to other sources.  In particular, one would expect a dispersion in the
observed quasar (QSO) spectra.  An analysis based on the first data
release of the Sloan Digital Sky Survey excludes a large part of the
parameter space~\cite{Ostman:2004eh}, suggesting that only for
$n_e\alt10^{-10}$~cm$^{-3}$ the axion mechanism is still able to
explain a dimming by $0.1$ magnitudes or more. If the QSO spectra had
an intrinsic dispersion at the 5\% level would rule out axion dimming
exceeding $\sim$~0.05~mag.  Future data will be sensitive to yet
larger regions in parameter space, yet QSOs will never be sensitive to
very low $n_e$.

A similar bound has been obtained by a possible violation of the
reciprocity relation between the luminosity distance and the
angular-diameter distance~\cite{bakuI,bakuII}. However, this
constraint is less robust than the QSO one because it is affected by
possibly large systematic errors that are difficult to
quantify~\cite{uam}.

The purpose of our paper is to further constrain the photon-axion
conversion model by studying its effect on the CMB spectral shape.  We
will show that the low-$n_e$ region of parameters left open by the QSO
limit is ruled out by our new limit, leaving little if any room for
the axion hypothesis to mimic cosmic acceleration.

In Sec.~\ref{Formalism} we discuss the formalism of photon-axion
conversion and in Sec.~\ref{dimming} we summarize its effect on SN~Ia
dimming.  In Sec.~\ref{CMBC} we describe the constraints coming from
spectral CMB distortions and in Sec.~\ref{QSO} we combine our new
limits with those from QSO dispersion. Finally, in
Sec.~\ref{conclusions} we draw our conclusions and comment on the
viability of the photon-axion conversion mechanism.

%%%%%%%%%%%%%%%%%%%%%%%%%%%%%%%%%%%%%%%%%%%%%%%%%%%%%%%%%%%%%%%%%%%%%%
\section{Photon-axion conversion}                    \label{Formalism}
%%%%%%%%%%%%%%%%%%%%%%%%%%%%%%%%%%%%%%%%%%%%%%%%%%%%%%%%%%%%%%%%%%%%%%

Axions and photons oscillate into each other in an external magnetic
field~\cite{sikivie,Raffelt:1987im,Anselm:1987vj,Raffeltbook} due to
the interaction term
\begin{equation}
{\cal L}_{a\gamma}=-\frac{1}{4}\,\gag
F_{\mu\nu}\tilde{F}^{\mu\nu}a=\gag{\bf E}\cdot{\bf B}\,a\,,
\end{equation}
where $F_{\mu\nu}$ is the electromagnetic field tensor,
$\tilde{F}_{\mu\nu}
=\frac{1}{2}\epsilon_{\mu\nu\rho\sigma}F^{\rho\sigma}$ is its dual,
$a$ is the axion field, and $\gag$ is the axion-photon coupling (with
dimension of inverse energy). We always use natural units with
$\hbar=c=k_{\rm B}=1$.  For very relativistic axions, the equations of
motion in the presence of an external magnetic field $B$ reduce to the
linearized form~\cite{Raffelt:1987im}
\begin{equation}\label{linsys1}
\left(\omega-i\partial_z +{\cal M}\right)  
\left(  
\begin{array}{ccc}  
A_x\\
A_y\\
a 
\end{array}  
\right)=0\,,  
\end{equation}  
where $z$ is the direction of propagation, $A_x$ and $A_y$ correspond
to the two linear polarization states of the photon field, and
$\omega$ is the photon or axion energy.  The mixing matrix is
\begin{equation}  \label{mixmatxy}
{\cal M}=\left(  
  \begin{array}{ccc}  
    \Delta_{xx}&\Delta_{xy}&\frac{1}{2}\gag B_x\\  
    \Delta_{yx}&\Delta_{yy}&\frac{1}{2}\gag B_y\\  
    \frac{1}{2}\gag B_x&\frac{1}{2}\gag B_y& \Delta_a 
  \end{array}  
  \right)\,,
\end{equation}
where $\Delta_a = -m^2_{a}/2\omega$. The component of ${\bf B}$
parallel to the direction of motion does not induce photon-axion
mixing.  The quantities $\Delta_{ij}$ with $i,j=x,y$ mix the photon
polarization states.  They are energy dependent and are determined
both by the properties of the medium and the QED vacuum polarization
effect.  We ignore the latter, being sub-dominant for the problem at
hand~\cite{Deffayet:2001pc}.

For a homogeneous magnetic field we may choose a coordinate system
aligned with the field direction. The linear photon polarization state
parallel to the transverse field direction ${\bf B}_T$ is denoted as
$A_\parallel$ and the orthogonal one as $A_\perp$.
Equation~(\ref{linsys1}) becomes then
\begin{equation}  \label{linsys}
\left(\omega-i\partial_z +{\cal M}\right)  
\left(  
\begin{array}{ccc}  
A_{\perp}\\
A_{\parallel}\\
a 
\end{array}  
\right)=0\,,
\end{equation}  
with mixing matrix  
\begin{equation}  \label{mixmat}
{\cal M}=\left(  
\begin{array}{ccc}  
\Delta_{\perp} &\Delta_{\rm R}     & 0\\ 
\Delta_R       &\Delta_{\parallel} & \Delta_{a\gamma}\\
0              &\Delta_{a\gamma}   & \Delta_a\\
  \end{array}  
  \right)\,.
\end{equation}
Here, $\Delta_{\perp}=\Delta_{\rm pl}+\Delta_{\perp}^{\rm CM}$,
$\Delta_{\parallel}=\Delta_{\rm pl}+\Delta_{\parallel}^{\rm CM}$,
$\Delta_{\rm pl}=-\omega_{\rm pl}^2/2\omega$, $\Delta_{a \gamma}=\gag
|{\bf B}_T|/2$, and $\omega_{\rm pl}^2 = 4\pi\alpha\,n_e/m_e$ defines
the plasma frequency, $m_e$ being the electron mass and $\alpha$ the
fine-structure constant.  The $\Delta^{\rm CM}_{\parallel, \perp}$
terms describe the Cotton-Mouton effect, i.e.\ the birefringence of
fluids in the presence of a transverse magnetic field where
$|\Delta_{\parallel}^{\rm CM}-\Delta_{\perp}^{\rm CM}|\propto B_T^2$.
These terms are of little importance for the following arguments and
will thus be neglected.  The Faraday rotation term $\Delta_{\rm R}$,
which depends on the energy and the longitudinal component $B_z$,
couples the modes $A_{\parallel}$ and $A_{\perp}$.  While Faraday
rotation is important when analyzing polarized sources of photons, it
plays no role for the problem at hand.

With this simplification the $A_\perp$ component decouples, and the
propagation equations reduce to a 2-dimensional mixing problem with a
purely transverse field ${\bf B}={\bf B}_T$
\begin{equation}
\left(\omega-i\partial_z +{\cal M}_2\right)  
\left(  
\begin{array}{cc}  
A_\parallel\\a 
\end{array}  
\right)=0,  
\end{equation}
with a 2-dimensional mixing matrix  
\begin{equation}  \label{mixmat2}
{\cal M}_{2}=\left(  
  \begin{array}{cc}  
    \Delta_{\rm pl}&\Delta_{a \gamma}\\  
    \Delta_{a \gamma}&\Delta_a  
  \end{array}  
  \right).
\end{equation} 
The solution follows from diagonalization through the rotation angle
\begin{equation}
\label{tan}
\vartheta = \frac{1}{2}\arctan
\left(\frac{2\Delta_{a \gamma}}{\Delta_{\rm pl}-\Delta_a}\right).
\end{equation}
In analogy to the neutrino case~\cite{Kuo:1989qe}, the probability for
a photon emitted in the state $A_{\parallel}$ to convert into an axion
after traveling a distance $s$ is
\begin{eqnarray}\label{p1ga}
  P_0(\gamma\rightarrow a)&=&
  \left|\langle A_\parallel(0)|a(s)\rangle\right|^2\nonumber\\
  &=&\sin^2\left(2 \vartheta \right)
  \sin^2(\Delta_{\rm osc}s/2)\nonumber\\  
  &=&\left(\Delta_{a \gamma} s\right)^2 
  \frac{\sin^2(\Delta_{\rm osc} s /2)}  
  {(\Delta_{\rm osc} s /2)^2} \;\ , 
\end{eqnarray}   
where the oscillation wavenumber is given by
\begin{equation}\label{deltaosc}
\Delta_{\rm osc}^2=(\Delta_{\rm pl}-\Delta_a)^2 +
4 \Delta_{a \gamma}^2\,.
\end{equation} 
The conversion probability is energy-independent when
$2|\Delta_{a\gamma}|\gg|\Delta_{\rm pl}-\Delta_{a}|$ or whenever the
oscillatory term in Eq.~(\ref{p1ga}) is small, i.e.\ $\Delta_{\rm osc}
s /2\ll1$, implying the limiting behavior $P_0=\left(\Delta_{a \gamma}
s\right)^2 \label{p1enind}$.

The propagation over many random $B$-field domains is a truly
3-dimensional problem, because different photon polarization states
play the role of $A_\parallel$ and $A_\perp$ in different
domains. This is enough to guarantee that the conversion probability
over many domains is an incoherent average over magnetic field
configurations and photon polarization states.  The probability after
travelling over a distance $r\gg s$, where $s$ is the domain size,
is~\cite{Grossman:2002by}
\begin{equation}\label{totpro}
P_{\gamma \to a}(r) = \frac{1}{3}
\left[1-\exp\left(-\frac{3P_0\,r}{2s}\right)\right]\,,
\end{equation}
with $P_0$ given by Eq.~(\ref{p1ga}).  As expected one finds that for
$r/s\to\infty$ the conversion probability saturates, so that on
average one third of all photons converts to axions.

%%%%%%%%%%%%%%%%%%%%%%%%%%%%%%%%%%%%%%%%%%%%%%%%%%%%%%%%%%%%%%%%%%%%%%
\section{Photon-axion conversion and Supernova dimming}
\label{dimming} 
%%%%%%%%%%%%%%%%%%%%%%%%%%%%%%%%%%%%%%%%%%%%%%%%%%%%%%%%%%%%%%%%%%%%%%

To explore the effect of photon-axion conversion on SN dimming we
recast the relevant physical quantities in terms of natural parameter
values.  The energy of optical photons is a few~eV. The strength of
widespread, all-pervading $B$-fields in the intergalactic medium must
be less than a few~$10^{-9}$~G over coherence lengths $s$ crudely at
the Mpc scale, according to the constraint
coming from the Faraday effect of distant radio
sources~\cite{Kronberg:1993vk}.  Along a given line of sight,
the number of such domains in our Hubble radius is about $N \approx
H_0^{-1}/s\approx 4 \times 10^3$ for $s\sim 1$~Mpc.  The mean diffuse
intergalactic plasma density is bounded by $n_e \lesssim 2.7 \times
10^{-7}$~cm$^{-3}$, corresponding to the recent WMAP measurement of
the baryon density~\cite{Spergel:2003cb}.  Recent results from the
CAST experiment~\cite{Andriamonje:2004hi} give a direct experimental
bound on the axion-photon coupling of $\gag \lesssim 1.16 \times
10^{-10}$~GeV$^{-1}$, comparable to the long-standing globular-cluster
limit~\cite{Raffeltbook}.  For ultra-light axions a stringent limit
from the absence of $\gamma$-rays from SN~1987A gives $\gag\lesssim
1\times 10^{-11}$~GeV$^{-1}$~\cite{Brockway:1996yr} or even $\gag
\lesssim 3\times 10^{-12}$~GeV$^{-1}$~\cite{Grifols:1996id}.
Therefore, suitable numerical values of the mixing parameters are
\begin{eqnarray}  \label{eq12nm}
\frac{\Delta_{a \gamma}}{{\rm Mpc}^{-1}} &=&
0.15\;g_{10}\;B_{\rm nG}
\,,\nonumber\\ 
\frac{\Delta_a} {{\rm Mpc}^{-1}} &=&
-7.7 \times 10^{28} \left(\frac{m_a}{1\,\rm eV}\right)^2
\left(\frac{\omega}{1\,\rm eV}\right)^{-1}
\,,\nonumber\\
\frac{\Delta_{\rm pl}}{{\rm Mpc}^{-1}}  &=& 
-11.1 \left(\frac{\omega}{1\,\rm eV}\right)^{-1}
\left(\frac{n_e}{10^{-7}\,{\rm cm}^{-3}}\right)\,, 
\end{eqnarray}
where we have introduced $g_{10}=\gag/10^{-10}$ GeV$^{-1}$ and
$B_{\rm nG}$ is the magnetic field strength in nano-Gauss.

The mixing angle defined in Eq.~(\ref{tan}) is too small to yield a
significant conversion effect for the allowed range of axion masses
because $|\Delta_a| \gg |\Delta_{a \gamma}| , |\Delta_{\rm pl}|$.
Therefore, to ensure a sufficiently large mixing angle one has to
require nearly massless pseudo-scalars, sometimes referred to as
``arions''~\cite{{Anselm:1981aw},{Anselm:1982ip}}.  Henceforth we will
consider the pseudoscalars to be effectively massless, so that our
remaining independent parameters are $g_{10}B_{\rm nG}$ and $n_e$.
Note that $m_a$ only enters the equations via the term $m_a^2 -
\omega_{\rm pl}^2$, so that for tiny but non-vanishing values of
$m_a$, the electron density should be interpreted as $n_{e,{\rm
eff}}=|n_e - m_a^2 m_e/(4 \pi \alpha)|$.

The distance relevant for SN~Ia dimming is the luminosity distance
$d_L$ at redshift $z$, defined by
\begin{equation}
\label{distance}
d^2_L(z) = \frac{\mathcal{L}}{4 \pi \mathcal{F}}\,,
\end{equation}
where $\mathcal{L}$ is the absolute luminosity of the source and
$\mathcal{F}$ is the energy flux arriving at
Earth~\cite{supnovR,supnovP}.
Usually the data are expressed in terms of
 magnitudes
\begin{equation}
m = M +  5 \log_{10} \left(\frac{d_L}{\rm  Mpc}\right)  + 25 \,,
\end{equation}
where $M$ is the absolute magnitude, equal to the value that $m$
would have at $d_L=10$ pc. 
After a distance $r$, photon-axion conversion has
reduced the number of photons emitted by the source and thus the flux
$\mathcal{F}$ to the fraction $P_{\gamma\to\gamma}=1-P_{\gamma\to a}$.
Therefore, the luminosity distance becomes
\begin{equation} 
d_L \to d_L/(P_{\gamma \to \gamma})^{1/2}
\end{equation}
and the brightness
\begin{equation}\label{isodim}
m \to m - \frac{5}{2}\log_{10}(P_{\gamma \to \gamma})\,.
\end{equation}
Distant SNe~Ia would eventually saturate ($P_{\gamma \to
\gamma}=2/3$), and hence they would appear $(3/2)^{1/2}$ times farther
away than they really are.  This corresponds to a maximum dimming of
approximately 0.4~mag.

\begin{figure}[t]
\centering
\epsfig{figure=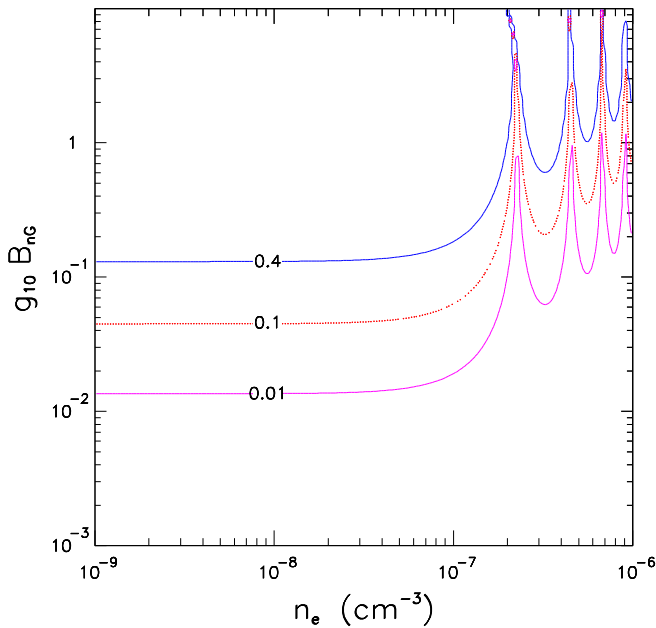,width =1.0\columnwidth,angle=0}
 \caption{\label{fig1}
   Iso-dimming curves for an attenuation of 0.01, 0.1, and
   0.4~magnitudes. The photon energy of $4.0$~eV is representative of
   the B-band. The size of a magnetic domain is $s=1$~Mpc.}
\end{figure}

In Fig.~\ref{fig1} we show qualitatively the regions of $n_e$ and
$g_{10}B_{\rm nG}$ relevant for SN dimming at cosmological distances.
To this end we show iso-dimming contours obtained from
Eq.~(\ref{isodim}) for a photon energy 4.0~eV and a magnetic domain
size $s=1$~Mpc.  For simplicity we neglect the redshift evolution of
the intergalactic magnetic field $B$, domain size $s$, plasma density
$n_e$ and photon frequency $\omega$. Our 
iso-dimming curves are intended  to illustrate  the  regions where the
photon-axion conversion could be relevant. In reality, the dimming should
be a more complicated function since the intergalactic medium is 
expected to be very irregular: there could be voids of  low $n_e$ density,
but there will also be high density clumps, 
sheets and filaments and these will typically have higher $B$
fields as well. However, the simplifications used in 
this work are consistent with the ones adopted in CKT II model and  
do not alter our main results.

%%%%%%%%%%%%%%%%%%%%%%%%%%% FIGURE 1 %%%%%%%%%%%%%%%%%%%%%%%%%%%%%%%%%
%%%%%%%%%%%%%%%%%%%%%%%%%%%%%%%%%%%%%%%%%%%%%%%%%%%%%%%%%%%%%%%%%%%%%%

The iso-dimming contours are horizontal in the low-$n_e$ and
low-$g_{10}B_{\rm nG}$ region.  They are horizontal for any
$g_{10}B_{\rm nG}$ when $n_e$ is sufficiently low.  From the
discussion in Sec.~\ref{Formalism} we know that the single-domain
probability $P_0$ of Eq.~(\ref{p1ga}) is indeed energy independent
when $|\Delta_{\rm osc} s| \ll 1$, i.e.\ for $|\Delta_{\rm pl}| s/2\ll
1$ and $|\Delta_{a \gamma}| s \ll 1$.  When $n_e\lesssim {\rm
few}~10^{-8}$~cm$^{-3}$ and $g_{10}B_{\rm nG} \lesssim 4$, we do not
expect an oscillatory behavior of the probability. This feature is
nicely reproduced in our iso-dimming contours.  From Fig.~\ref{fig1}
we also deduce that a significant amount of dimming is possible only
for $g_{10}B_{\rm nG}\gtrsim 4\times 10^{-2}$.

In CKT~I, where the effect of $n_e$ was neglected, a value $m_a \sim
10^{-16}$ eV was used.  In terms of our variables, this corresponds to
$n_{e,\rm eff} \approx 6 \times 10^{-12}$~cm$^{-3}$. As noted in
CKT~II, when plasma effects are taken into account, any value
$n_e\lesssim 2.5 \times 10^{-8}$~cm$^{-3}$ guarantees the required
achromaticity of the dimming below the 3\% level between the B and V
bands.  The choice $B_{\rm nG}$ of a few and $g_{10}\approx 0.1$ in
CKT~I and~II falls in the region where the observed SN dimming could
be explained while being marginally compatible with the bounds on $B$
and $g_{10}$.

%%%%%%%%%%%%%%%%%%%%%%%%%%%%%%%%%%%%%%%%%%%%%%%%%%%%%%%%%%%%%%%%%%%%%%
\section{CMB Constraints}                                 \label{CMBC}
%%%%%%%%%%%%%%%%%%%%%%%%%%%%%%%%%%%%%%%%%%%%%%%%%%%%%%%%%%%%%%%%%%%%%%

If $\gamma\to a$ conversion over cosmological distances is responsible
for the SN~Ia dimming, the same phenomenon should also leave an
imprint in the CMB.  We note that a similar argument was previously
considered for photon$\to{}$graviton conversion~\cite{Chen:1994ch}.
Qualitatively, in the energy-dependent region of $P_{\gamma\to a}$ one
expects a rather small effect due to the low energy of CMB photons
($\omega \sim 10^{-4}$~eV). However, when accounting for the
incoherent integration over many domains crossed by the photon,
appreciable spectral distortions may arise in view of the accuracy of
the CMB data (at the level of one part in $10^{4}$--$10^{5}$). For the
same reason, in the energy-independent region, at much lower values of
$n_e$ than for the SNe~Ia, the constraints on $g_{10}B_{\rm nG}$ are
expected to be quite severe.  The depletion of CMB photons in the
patchy magnetic sky and its effect on the CMB anisotropy pattern have
been previously considered in~\cite{Csaki:2001yk}.  However, more
stringent limits come from the distortion of the overall blackbody
spectrum.

To this end we use the COBE/FIRAS data for the experimentally measured
spectrum, corrected for foregrounds~\cite{Fixsen:1996nj}.  Note that
the new calibration of FIRAS~\cite{Mather:1998gm} is within the old
errors and would not change any of our conclusions.  The $N = 43$ data
points $\Phi^{\rm exp}_i$ at different energies $\omega_i$ are
obtained by summing the best-fit blackbody spectrum  to the
residuals reported in Ref.~\cite{Fixsen:1996nj}.  The experimental
errors $\sigma_i$ and the  correlation indices
$\rho_{ij}$ between different energies are also available.  In the presence of
photon-axion conversion, the original intensity of the ``theoretical
blackbody'' at temperature $T$
\begin{equation}
\label{planck}
\Phi^0({\omega},T) = \frac{\omega^3}{ 2 \pi^2} 
\big[ \exp (\omega/T )-1 \big]^{-1}
\end{equation}
would convert to a deformed spectrum that is given by
$\Phi({\omega},T)=\Phi^0({\omega},T)P_{\gamma\to\gamma}({\omega})$.
We then build the reduced chi-squared function
\begin{equation}
\chi_\nu^2(T,\lambda)=\frac{1}{{N}-1}
\sum_{i,j=1}^{N} {\Delta \Phi_i} (\sigma^2)^{-1}_{ij}
{\Delta \Phi_j} \,,
\end{equation}
where 
\begin{equation}
\Delta \Phi_i = \Phi^{\rm exp}_i-\Phi^0({\omega}_i,T)
P_{\gamma\to\gamma}({\omega_i},\lambda)
\end{equation}
is the $i$-th residual, and
\begin{equation}
\sigma^2_{ij}= \rho_{ij} \sigma_{i} \sigma_{j} 
\end{equation}
is the covariance matrix.
We minimize this function with respect to $T$ 
%%%%%%%%%%%%%%%%%%%%%%%%%%%%%%%%%%%%%%%%%%%%%%%%%%%%%%%%%%%%%%%%%
\footnote{ In principle, one should  marginalize also over the galactic foreground 
 spectrum~\cite{Fixsen:1996nj}. However, we neglect it
 since it is a subleading effect with respect
to the deformation induced on the CMB blackbody  by the photon-axion conversion.}
%%%%%%%%%%%%%%%%%%%%%%%%%%%%%%%%%%%%%%%%%%%%%%%%%%%%%%%%%%%%%%%%%
for each point in the
parameter space $\lambda=(n_e,g_{10}B_{\rm nG})$, i.e.\ $T$ is an
empirical parameter determined by the $\chi_\nu^2$ minimization for
each $\lambda$ rather than being fixed at the standard value
$T_0=2.725\pm0.002$~K \cite{Mather:1998gm}.

%%%%%%%%%%%%%%%%%%%%%%%%%%% FIGURE 2 %%%%%%%%%%%%%%%%%%%%%%%%%%%%%%%%%
\begin{figure}[t]
\centering
\epsfig{figure=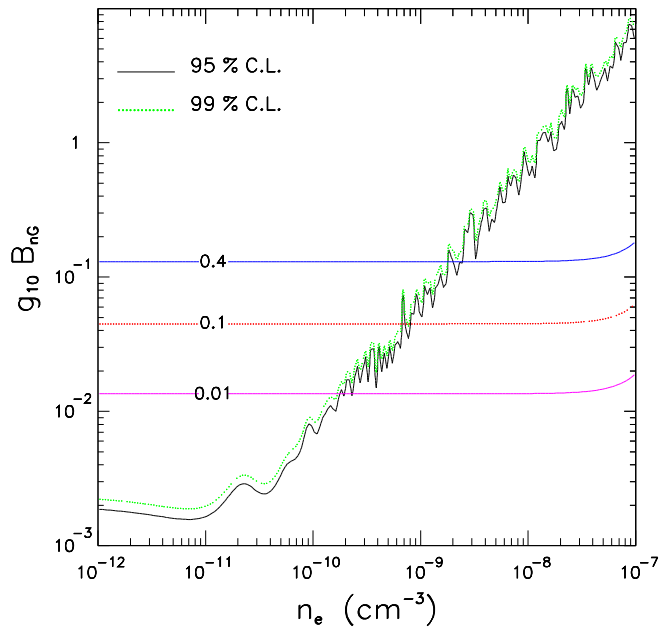, width =1.0\columnwidth, angle=0}
\caption{\label{fig2}Exclusion plot for axion-photon conversion based
  on the COBE/FIRAS CMB spectral data.  The region above the solid
  curve is excluded at 95\% C.L. whereas the one above the dotted
  curve is excluded at 99\% C.L. The size of each magnetic domain is
  fixed at $s=1$~Mpc. We also reproduce the iso-dimming contours from
  Fig.~\ref{fig1}.}
\end{figure}
%%%%%%%%%%%%%%%%%%%%%%%%%%%%%%%%%%%%%%%%%%%%%%%%%%%%%%%%%%%%%%%%%%%%%%

In Fig.~\ref{fig2} we show our exclusion contour in the plane of $n_e$
and $g_{10}B_{\rm nG}$. The region above the continuous curve is the
excluded region at 95\% C.L., i.e.\ in this region the chance
probability to get larger values of $\chi_\nu^2$ is lower than~5\%. We
also show the corresponding 99\% C.L. contour which is very close to
the 95\% contour so that another regression method and/or exclusion
criterion would not change the results very much.  Within a factor of
a few, the same contours also hold if one varies the domain size $s$
within a factor~10.

Comparing our exclusion plot with the iso-dimming curves of
Fig.~\ref{fig1} we conclude that the entire region $n_e \lesssim
10^{-9}$~cm$^{-3}$ is excluded for SN dimming.

A few comments are in order.  Intergalactic magnetic fields probably
are a relatively recent phenomenon in the cosmic history, arising only
at redshifts of a few. As a first approximation we have then
considered the photon-axion conversion as happening on present ($z=0$)
CMB photons.  Since $P_{\gamma\to \gamma}$ is an increasing function
of the photon energy $\omega$, our approach leads to conservative
limits.  Moreover, we assumed no correlation between $n_e$ and the
intergalactic magnetic field strength. It is however physically
expected that the fields are positively correlated with the plasma
density so that relatively high values of $g_{10}B_{\rm nG}$ should be
more likely when $n_e$ is larger.  Our constraints in the region of
$n_e\gtrsim 10^{-10}$~cm$^{-3}$ are thus probably tighter than what
naively appears.

%%%%%%%%%%%%%%%%%%%%%%%%%%%%%%%%%%%%%%%%%%%%%%%%%%%%%%%%%%%%%%%%%%%%%%
\section{QSO Constraints}                                  \label{QSO}
%%%%%%%%%%%%%%%%%%%%%%%%%%%%%%%%%%%%%%%%%%%%%%%%%%%%%%%%%%%%%%%%%%%%%%

\begin{figure}[ht]
\centering
\epsfig{figure=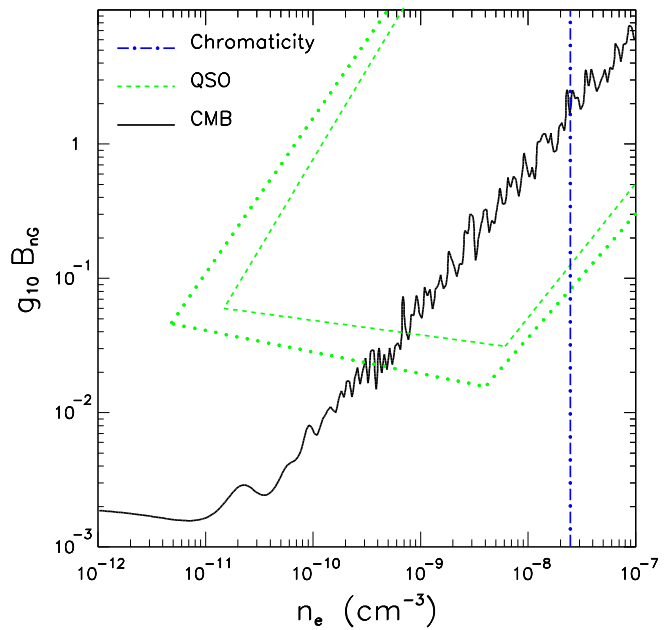, width =1.0\columnwidth, angle=0}
 \caption{\label{fig3}Exclusion plot for photon-axion conversion.  The
   region to the right of the dot-dashed line is excluded by requiring
   achromaticity of SN~Ia dimming. The region inside the dashed lines
   is excluded by the dispersion in QSO spectra. Assuming an intrinsic
   dispersion of 5\% in QSO spectra, the excluded region could be
   extended up to the dotted curve. Our CMB argument excludes the
   entire region above the continuous curve at 95\% C.L.}
\end{figure}

Our limits are nicely complementary to the ones obtained from the
effects of photon-axion conversion on quasar colors and
spectra~\cite{Ostman:2004eh}.  In Fig.~\ref{fig3} we superimpose our
CMB exclusion contours with the schematic region excluded by quasars
%%%%%%%%%%%%%%%%%%%%%%%%%%%%%%%%%%%%%%%%%%%%%%%%%%%%%%%%%%%%%%%%%%%%%%
\footnote{We use the exclusion regions of astro-ph/0410501v1.
  In the published version~\cite{Ostman:2004eh}, corresponding to
  astro-ph/0410501v2, the iso-dimming curves were erroneously changed.
  The difference is that in version~1 the angle $\alpha$ in Eq.~(3) of
  Ref.~\cite{Ostman:2004eh} that characterizes the random magnetic
  field direction was correctly taken in the interval
  0--360$^\circ$ whereas in version 2 it was taken in the interval
  0--90$^\circ$ (private communication by the authors).}.
%%%%%%%%%%%%%%%%%%%%%%%%%%%%%%%%%%%%%%%%%%%%%%%%%%%%%%%%%%%%%%%%%%%%%%
The region to the right of the dot-dashed line is excluded by
requiring achromaticity of SN~Ia dimming~\cite{Csaki:2001jk}. The
region inside the dashed lines is excluded by the dispersion in QSO
spectra. Moreover, assuming an intrinsic dispersion of 5\% in these
spectra, the excluded region could be enlarged up to the dotted
lines. Our CMB argument excludes the region above the solid curve at
95\% C.L. 

A cautionary remark is in order when combining the two constraints.
As we have discussed in the previous section, 
our CMB limits on photon-axion conversion are model independent.
 Conversely,  
the limits placed by the QSO spectra are possibly subjected
to loop holes, since they are based on a full
  correlation
between  the intergalactic electron density and the magnetic field strength, 
which is reasonable but not well established observationally.

%%%%%%%%%%%%%%%%%%%%%%%%%%% FIGURE 3 %%%%%%%%%%%%%%%%%%%%%%%%%%%%%%%%%
%%%%%%%%%%%%%%%%%%%%%%%%%%%%%%%%%%%%%%%%%%%%%%%%%%%%%%%%%%%%%%%%%%%%% 

%%%%%%%%%%%%%%%%%%%%%%%%%%%%%%%%%%%%%%%%%%%%%%%%%%%%%%%%%%%%%%%%%%%%%%
\section{Conclusions}                              \label{conclusions}
%%%%%%%%%%%%%%%%%%%%%%%%%%%%%%%%%%%%%%%%%%%%%%%%%%%%%%%%%%%%%%%%%%%%%%

We have examined the conversion of CMB photons into very low-mass
axions in the presence of intergalactic magnetic fields.  The
resulting CMB spectral deformation excludes a previously allowed
parameter region corresponding to very low densities of the
intergalactic medium. Our new limits are complementary to the ones
derived from QSO dispersion which place serious constraints on the
axion-photon conversion mechanism.  As a result, it appears that
this mechanism can hardly play a leading
role for the apparent SN~Ia dimming.

The axion-photon conversion hypothesis has also been advocated to
explain trans-GZK cutoff events in Ultra High Energy Cosmic Rays
(UHECRs)~\cite{Csaki:2003ef}.  In principle, UHECR photons, produced
in cosmological sources far away, could drastically reduce energy
losses while propagating in the intergalactic medium as axions.  Some
of these particles would eventually convert back to photons within a
few GZK radii, thus justifying the observations of extremely high
energy events as well as their isotropy.  While one can not rule out
the possibility that some UHE ``photon-like'' events at energies
$E\gtrsim 4\times 10^{19}$~eV might be due to this mechanism, our
bounds imply that it can play only a subdominant role.  Moreover,
photons anyway are disfavored as candidates for the majority of the
UHECRs.

In summary, the CMB constraints together with previous limits suggest
that the fascinating mechanism of photon-axion conversion in the
intergalactic magnetic fields does not play an important role for
either the phenomenon of SN~Ia dimming or for UHECR propagation.  A
definitive verdict would probably require a common analysis of SN~Ia
dimming, QSO spectra, and the Faraday effect of distant radio sources,
based on mutually consistent assumptions about the intergalactic
matter density and its distribution, the intergalactic $B$-field
strength and its distribution and correlation with the electron
density, and the redshift evolution of these quantities. Our results
show that the low-$n_e$ escape route from the QSO limits is definitely
closed.

%%%%%%%%%%%%%%%%%%%%%%%%%%%%%%%%%%%%%%%%%%%%%%%%%%%%%%%%%%%%%%%%%%%%%%
%% Acknowledgments %%%%%%%%%%%%%%%%%%%%%%%%%%%%%%%%%%%%%%%%%%%%%%%%%%%
%%%%%%%%%%%%%%%%%%%%%%%%%%%%%%%%%%%%%%%%%%%%%%%%%%%%%%%%%%%%%%%%%%%%%%
\begin{acknowledgments}
We thank S.~Hannestad, M.~Kachelrie\ss$\:$ and T.~Rashba for useful discussions
and L.~Ostman and E.~M\"ortsell for clarifying some issues about their
quasar bounds~\cite{Ostman:2004eh}. A.M.~thanks G.L.~Fogli for
interesting discussions, and E.~Lisi and D.~Montanino for carefully
reading the manuscript. We thank C.~Cs\'aki, N.~Kaloper, and
J.~Terning for critical comments on an earlier version of this
manuscript.  We acknowledge partial support by the Deutsche
Forschungsgemeinschaft under Grant No.~SFB-375 and by the European
Union under the Ilias project, contract No.~RII3-CT-2004-506222.  The
work of A.M. is supported in part by the Italian ``Istituto Nazionale
di Fisica Nucleare'' (INFN) and by the ``Ministero dell'Istruzione,
Universit\`a e Ricerca'' (MIUR) through the ``Astroparticle Physics''
research project.
\end{acknowledgments}

%\clearpage

%%%%%%%%%%%%%%%%%%%%%%%%%%%%%%%%%%%%%%%%%%%%%%%%%%%%%%%%%%%%%%%%%%%%%%
%%%  Bibliography  %%%%%%%%%%%%%%%%%%%%%%%%%%%%%%%%%%%%%%%%%%%%%%%%%%%
%%%%%%%%%%%%%%%%%%%%%%%%%%%%%%%%%%%%%%%%%%%%%%%%%%%%%%%%%%%%%%%%%%%%%%

%%%%%%%%%%%%%%%%%%%%%%%%%%%%%%%%%%%%%%%%%%%%%%%%%%%%%%%%%%%%%%%%%%%%%%

\begin{thebibliography}{00}

\bibitem{supnovR}
  A.~G.~Riess {\it et al.} (Supernova Search Team Collaboration),
  ``Observational evidence from supernovae for an accelerating
  universe and a cosmological constant,''
  Astron.\ J.\  {\bf 116}, 1009 (1998)
  [astro-ph/9805201].

\bibitem{supnovP}
  S.~Perlmutter {\it et al.}
  (Supernova Cosmology Project Collaboration),
  ``Measurements of $\Omega$ and $\Lambda$ from 42 high-redshift
  supernovae,''
  Astrophys.\ J.\  {\bf 517}, 565 (1999)
  [astro-ph/9812133].

\bibitem{Riess:2004nr}
  A.~G.~Riess {\it et al.}  (Supernova Search Team Collaboration),
  ``Type Ia supernova discoveries at $z>1$ from the Hubble Space
  Telescope: Evidence for past deceleration and constraints on
  dark energy evolution,''
  Astrophys.\ J.\  {\bf 607}, 665 (2004)
  [astro-ph/0402512].
  %%CITATION = ASTRO-PH 0402512;%%

\bibitem{Spergel:2003cb}
  D.~N.~Spergel {\it et al.}  [WMAP Collaboration],
  ``First Year Wilkinson Microwave Anisotropy Probe (WMAP)
  observations: Determination of cosmological parameters,''
  Astrophys.\ J.\ Suppl.\  {\bf 148}, 175 (2003)
  [astro-ph/0302209].
  %%CITATION = ASTRO-PH 0302209;%%

\bibitem{Tegmark:2003ud}
  M.~Tegmark {\it et al.}  [SDSS Collaboration],
  ``Cosmological parameters from SDSS and WMAP,''
  Phys.\ Rev.\ D {\bf 69}, 103501 (2004)
  [astro-ph/0310723].
  %%CITATION = ASTRO-PH 0310723;%%

\bibitem{Carroll:2003qq}
  S.~M.~Carroll,
  ``Why is the universe accelerating?,''
  eConf {\bf C0307282} (2003) TTH09
  [AIP Conf.\ Proc.\  {\bf 743}, 16 (2005),
  astro-ph/0310342].

\bibitem{Csaki:2001yk}
  C.~Cs\'aki, N.~Kaloper and J.~Terning (CKT~I)
  ``Dimming supernovae without cosmic acceleration,''
  Phys.\ Rev.\ Lett.\  {\bf 88}, 161302 (2002)
  [hep-ph/0111311].

\bibitem{Deffayet:2001pc}
  C.~Deffayet, D.~Harari, J.~P.~Uzan and M.~Zaldarriaga,
  ``Dimming of supernovae by photon-pseudoscalar conversion and the
  intergalactic plasma,''
  Phys.\ Rev.\ D {\bf 66}, 043517 (2002) 
  [hep-ph/0112118].

\bibitem{Csaki:2001jk}
  C.~Cs\'aki, N.~Kaloper and J.~Terning (CKT~II),
  ``Effects of the intergalactic plasma on supernova dimming via
  photon axion oscillations,''
  Phys.\ Lett.\ B {\bf 535}, 33 (2002)
  [hep-ph/0112212].

\bibitem{Ostman:2004eh}
  L.~Ostman and E.~M\"ortsell,
  ``Limiting the dimming of distant type Ia supernovae,''
  JCAP {\bf 0502}, 005 (2005)
  [astro-ph/0410501].
  
\bibitem{bakuI}
  B.~A.~Bassett and M.~Kunz,
  ``Cosmic acceleration versus axion photon mixing,''
  Astrophys.\ J.\  {\bf 607}, 661 (2004)
  [astro-ph/0311495].
  %%CITATION = ASTRO-PH 0311495;%%

\bibitem{bakuII}
  B.~A.~Bassett and M.~Kunz,
  ``Cosmic distance-duality as a probe of exotic physics and
  acceleration,''
  Phys.\ Rev.\ D {\bf 69}, 101305 (2004) 
  [astro-ph/0312443].
  %%CITATION = ASTRO-PH 0312443;%%

\bibitem{uam}
  J.~P.~Uzan, N.~Aghanim and Y.~Mellier,
  ``The distance duality relation from x-ray and
  Sunyaev-Zel'dovich observations of clusters,''
  Phys.\ Rev.\ D {\bf 70}, 083533 (2004) 
  [astro-ph/0405620]. 
  %%CITATION = ASTRO-PH 0405620;%%

\bibitem{sikivie}
  P.~Sikivie,
  ``Experimental tests of the ``invisible'' axion,''
  Phys.\ Rev.\ Lett.\  {\bf 51}, 1415 (1983), 
  Erratum {\it ibid.} {\bf 52}, 695 (1984).
  %%CITATION = PRLTA,51,1415;%%

\bibitem{Raffelt:1987im}
  G.~Raffelt and L.~Stodolsky,
  ``Mixing of the photon with low mass particles,''
  Phys.\ Rev.\ D {\bf 37}, 1237 (1988).
  %%CITATION = PHRVA,D37,1237;%%

\bibitem{Anselm:1987vj}
  A.~A.~Anselm,
  ``Experimental test for arion $\leftrightarrow$ photon 
  oscillations in a homogeneous constant magnetic field,''
  Phys.\ Rev.\ D {\bf 37}, 2001 (1988).
  
\bibitem{Raffeltbook}
	 G.~G.~Raffelt,
  ``Particle physics from stars,''
  Ann.\ Rev.\ Nucl.\ Part.\ Sci.\  {\bf 49}, 163 (1999)
  [hep-ph/9903472];
  G.~G.~Raffelt,
  {\it Stars as Laboratories for Fundamental Phys\-ics},
  (Univ.\ of Chicago Press, 1996).

\bibitem{Kuo:1989qe}
  T.~K.~Kuo and J.~T.~Pantaleone,
  ``Neutrino oscillations in matter,''
  Rev.\ Mod.\ Phys.\  {\bf 61}, 937 (1989).
  %%CITATION = RMPHA,61,937;%%

\bibitem{Grossman:2002by}
  Y.~Grossman, S.~Roy and J.~Zupan,
  ``Effects of initial axion production and photon axion
  oscillation on type Ia supernova dimming,''
  Phys.\ Lett.\ B {\bf 543}, 23 (2002)
  [hep-ph/0204216].
  %%CITATION = HEP-PH 0204216;%%

\bibitem{Kronberg:1993vk}
  P.~P.~Kronberg,
  ``Extragalactic magnetic fields,''
  Rept.\ Prog.\ Phys.\  {\bf 57}, 325 (1994).
  %%CITATION = RPPHA,57,325;%%

\bibitem{Andriamonje:2004hi}
  K.~Zioutas {\it et al.}  (CAST Collaboration),
  ``First results from the CERN axion solar telescope (CAST),''
  Phys.\ Rev.\ Lett.\  {\bf 94}, 121301 (2005)
  [hep-ex/0411033].
  %%CITATION = HEP-EX 0411033;%%

\bibitem{Brockway:1996yr}
  J.~W.~Brockway, E.~D.~Carlson and G.~G.~Raffelt,
  ``SN 1987A gamma-ray limits on the conversion of pseudoscalars,''
  Phys.\ Lett.\ B {\bf 383}, 439 (1996)
  [astro-ph/ 9605197].

\bibitem{Grifols:1996id}
  J.~A.~Grifols, E.~Mass\'o and R.~Toldr\`a,
  ``Gamma rays from SN~1987A due to pseudoscalar conversion,''
  Phys.\ Rev.\ Lett.\ {\bf 77}, 2372 (1996)
  [astro-ph/9606028].
  %%CITATION = ASTRO-PH 9606028;%%

\bibitem{Anselm:1981aw}
  A.~A.~Anselm and N.~G.~Uraltsev,
  ``A second massless axion?,''
  Phys.\ Lett.\ B {\bf 114}, 39 (1982).
  %%CITATION = PHLTA,B114,39;%%

\bibitem{Anselm:1982ip}
  A.~A.~Anselm and N.~G.~Uraltsev,
  ``Long range `arion' field in the radiofrequency band,''
  Phys.\ Lett.\ B {\bf 116}, 161 (1982).
  %%CITATION = PHLTA,B116,161;%%
  
\bibitem{Chen:1994ch}
  P.~Chen,
  ``Resonant photon-graviton conversion and cosmic microwave
  background fluctuations,''
  Phys.\ Rev.\ Lett.\ {\bf 74}, 634 (1995); 
  Erratum {\it ibid.} {\bf 74}, 3091 (1995).

\bibitem{Fixsen:1996nj}
  D.~J.~Fixsen, E.~S.~Cheng, J.~M.~Gales, J.~C.~Mather, R.~A.~Shafer
  and E.~L.~Wright,
  ``The cosmic microwave background spectrum from the full 
  COBE/FIRAS data set,''
  Astrophys.\ J.\ {\bf 473}, 576 (1996)
  [astro-ph/9605054].

\bibitem{Mather:1998gm}
  J.~C.~Mather, D.~J.~Fixsen, R.~A.~Shafer, C.~Mosier
  and D.~T.~Wilkinson,
  ``Calibrator design for the COBE far infrared absolute
  spectrophotometer (FIRAS),''
  Astrophys.\ J.\  {\bf 512} (1999) 511
  [astro-ph/9810373].

\bibitem{Csaki:2003ef}
  C.~Cs\'aki, N.~Kaloper, M.~Peloso and J.~Terning,
  ``Super-GZK photons from photon axion mixing,''
  JCAP {\bf 0305}, 005 (2003)
  [hep-ph/0302030].
  %%CITATION = HEP-PH 0302030;%%

\end{thebibliography}
\end{document}